\documentclass{vgtc}                          

\ifpdf
  \pdfoutput=1\relax                   
  \usepackage{graphicx}                
  \DeclareGraphicsExtensions{.pdf,.png,.jpg,.jpeg} 
\else
  \usepackage{graphicx}                
  \DeclareGraphicsExtensions{.eps}     
\fi%

\graphicspath{{figures/}{pictures/}{images/}{./}} 

\usepackage{microtype}                 
\PassOptionsToPackage{warn}{textcomp}  
\usepackage{textcomp}                  
\usepackage{mathptmx}                  
\usepackage{times}                     
\usepackage{cite}                      
\usepackage{tabu}                      
\usepackage{booktabs}                  
\usepackage{appendix}
\usepackage{enumitem}

\onlineid{0}

\vgtccategory{Research}

\vgtcinsertpkg

\title{Replicability and Transparency for the Creation of Public Human User Video Game Datasets}

\author{Emma J. Pretty\thanks{e-mail: emma.pretty@rmit.edu.au}
\and Renan Guarese\thanks{e-mail: renan.guarese@rmit.edu.au}
\and Haytham M. Fayek\thanks{e-mail: haytham.fayek@ieee.org} 
\and Fabio Zambetta\thanks{e-mail: fabio.zambetta@rmit.edu.au}}
\affiliation{\scriptsize RMIT University, Australia}

\teaser{
  \centering
\includegraphics[keepaspectratio, height=6cm, width=9cm]{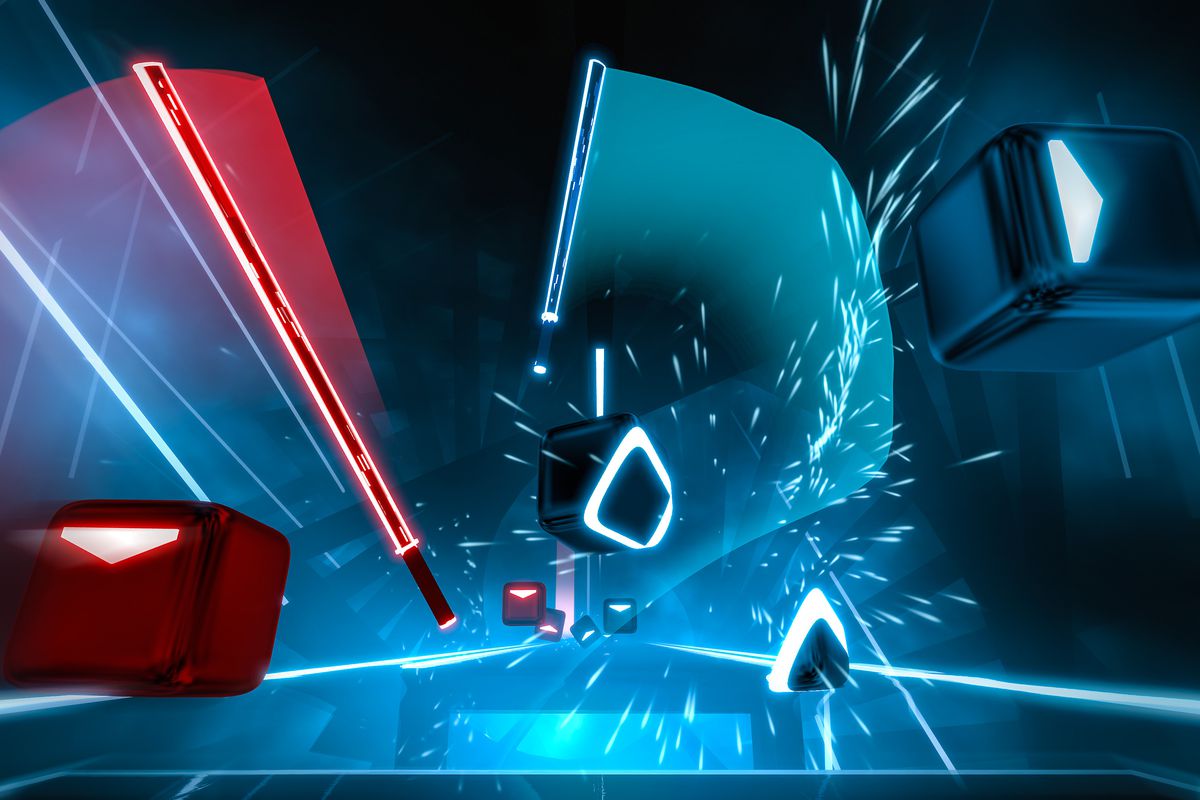}
  \caption{Beat Saber for VR by Beat Games~\cite{beatsaber_2019}}
  \label{fig:teaser}
}

\abstract{
Replicability is absent in games research; a lack of transparency in protocol detail hinders scientific consensus and willingness to publish public datasets, impacting the application of these techniques in video games research. To combat this, we propose and give an example of the use of a set of experimental considerations, such as games and materials choice. This work promotes the communication of research protocols when publishing datasets, benefiting researchers when designing experiments.

} 

\CCScatlist{
  \CCScatTwelve{Human-centered computing}{Collaborative and social computing}{Collaborative and social computing systems and tools}{Open source software};
  \CCScatTwelve{Applied computing}{Computers in other domains}{Personal computers and PC applications}{Computer games}
}

\begin{document}
\maketitle

\section{Introduction}

As with many fields of science, replicability is a large problem in human user video games research. Aside from researchers conducting the same experiment under the same or similar conditions again (known as repeatability~\cite{crandall2016scientific}), very little research explores the exact same experimental task, measurement tools, and analysis from another experiment, which is a powerful method of systematically corroborating findings for scientific consensus of a game element or construct. 

Take the following hypothetical scenario: two independent researchers explore the effects of 3 hours of sleep deprivation on self-reported cognitive load during video game play. The first researcher used the commercially available game Beat Saber~\cite{beatsaber_2019} for virtual reality (VR; see Figure~\ref{fig:teaser}) and a custom-made questionnaire regarding cognitive load, whereas the second researcher decides to use a custom-built VR rhythm game and measured cognitive load with the NASA Task Load Index (NASA-TLX)~\cite{hart1988development}. All other experimental methodologies and details were the same. \label{hypothetical}

According to its definition, this is an example of conceptual replication, as the second researcher aims to explore the same underlying concept or hypothesis as the first, but with different experimental designs~\cite{crandall2016scientific}. This does not guarantee, however, that the same outcomes are produced. There are a number of aspects that could impact this, however the main point of interest in this paper is the use of custom-built tasks and instruments to test these types of hypotheses. Custom instruments are not inherently harmful, however are often prone to experimenter bias and are unlikely to have their source code or original materials published for other researchers to see or use~\cite{mcmahan2011considerations}. If the first study in our hypothetical example found significantly higher self-reported cognitive load whereas the second had null results, how could we explain this difference? 

Aside from the more direct variables that could impact these results, such as self-report biases, prior knowledge of the game, and individualised impacts of sleep-deprivation, we believe that the measurement tools and experiment tasks themselves should be observed under a more critical eye, especially when attempting to replicate a study. Related to this is transparency; which in this case includes the complete documentation and dissemination of the methodology and procedure used, the source code for custom built tools or measurements, and the availability of the code used to process and analyse the data, which is also related to reproducibility~\cite{elliott2022taxonomy}. Replicability and transparency go hand-in-hand; with a lack of transparency there cannot be accurate direct or systematic replication of studies, and unsuccessful replication studies will often fail to be transparent or will not be published at all~\cite{national2019reproducibility}. This is problematic as a core pillar of the scientific approach, but also for the creation of public datasets, of which there are only a few in human user video games research. In this paper, we present some core considerations when developing experimental designs that encourage the publishing of datasets:
\begin{itemize} [noitemsep]
    \item We explain the factors to consider when choosing the tools that measure the variables of interest (Section~\ref{measure})
    \item We provide details on how to utilise games as an experimental stimulus in order to promote transparency and replicability (Section~\ref{game})
    \item A number of considerations for methodology are also listed in order to minimise variability within and between studies, thus strengthening future replication studies (Section~\ref{procedure})
    \item Lastly, the paper finishes with an example of how the discussed theoretical considerations can be used in practice with distinct rationale for each choice made (Section~\ref{casestudy}).
\end{itemize}

\section{Related Works}
\subsection{Games as Stimuli}
Many works have described some key criteria of how to choose and use video games as experimental tasks, highlighting the benefits and limitations of using off-the-shelf and custom games~\cite{jarvela2014practical,mcmahan2011considerations}. There is consensus that off-the-shelf games, whilst allowing better ecological validity (study generalisability), are complicated stimulus that might make it difficult to draw appropriate conclusions about a single aspect of gameplay. They are also often less accessible for modification due to not being open source. Custom games, on the other hand, are less refined, engaging, and natural, which is detrimental for external validity and making claims about player behaviour. They do benefit the researcher's ability to completely control the experimental environment, however. A key limitation of custom games is also experimenter bias - in creating the game, the researcher is aware of the construct under investigation and thus the experiment may be designed (intetionally or not) in a way that is more prone to producing the favoured results. 

Going beyond traditional computer or console games, the same considerations are valid for XR games. For instance, Yildrim \cite{yildirim2019cybersickness} chose two off-the-shelf games for their cybersickness comparisons on the basis that both titles had comparable desktop and VR versions, in both versions the player perspective could be changed with controller input or head movements, and both prod visually-induced self-motion while navigating through a fast-paced environment. Similarly, Shelstad et al. \cite{Shelstad2017gaming} used an off-the-shelf tower defense game with desktop and VR versions to assess user experience on a game satisfaction scale in both platforms. In both studies, already-established commercial games were chosen due to their features corresponding to their research needs. On the other hand, several custom VR games have been proposed, developed, and user-tested in academia for specific research purposes, commonly in an educational or serious game context \cite{oyelere2020exploring, checa2020review}. For the purpose of replicability and transparency however, both options are valid choices but do require extra considerations when considering publishing a dataset from the study, which we will cover later in this paper.

\subsection{Free access to data}
\label{subsecFree}
A number of reviews and articles have pushed for the mandatory availability of data underlying scientific publications as a fundamental of the scientific method~\cite{andreoli2014open}. Science is built upon the drawing of conclusions from data, and almost half of researchers will use data generated by other scientists at some point~\cite{tedersoo2021data}. Despite this, researchers across all scientific fields are not willing to share nor do they make their data available. For example, in an empirical study of 488 economics researchers, it was found that 80\% of researchers do not voluntarily share their data~\cite{andreoli2014open}. Another study~\cite{tedersoo2021data} requested access to data from 310 papers that included data availability statements (e.g. “contact author for data"), published in \textit{Nature} or \textit{Science}. They found that only 39.4\% of authors accepted the requests, with 19.4\% declining and 41.3\% not responding (60-day maximum response period). Despite the majority rejecting and ignoring these requests, it has been found that papers with links to data in a repository have up to a 25\% increase in number of citations~\cite{colavizza2020citation}. Whilst the publishing of data alongside its corresponding publications benefits science as a whole, this data is not always usable. This may be due to limitations of the publisher or the confidentiality requirements of the data, but a large barrier to a freely accessible dataset is its completeness and comprehensibility from poor annotation and documentation~\cite{molloy2011open}. Attempts have been made in the field of machine learning for the standardisation of published datasets to be used in applying AI algorithms. The authors of~\cite{gebru2021datasheets} provided a series of questions covering the entire scientific process for researchers to use when carefully reflecting on and designing a dataset and it's accompanying documentation. Similarly, the NeurIPS conference provides its authors with a checklist~\cite{neurips_checklist} of guidelines that address transparency, reproducibility, and ethics. This checklist has been used in the evaluation of research dissemination that incentivises the researcher to carefully consider these areas.

Specifically regarding VR user studies, Lanier et al. \cite{Lanier2019Virtual} performed a meta analysis of 61 academic papers regarding their data analysis transparency and validity. Their research exposed a high error rate in statistical reporting and an overall low transparency across the studies surveyed, with less than 4\% of those making their datasets publicly available.

\subsection{Human user datasets}
Regarding video game datasets, a large number exist for the purpose of applying AI algorithms (see~\cite{justesen2019deep} for an overview of some common games AI platforms and datasets), and are commonly made up of publicly available video footage of gameplay such as~\cite{lin2017stardata} and~\cite{guss2019minerl}, however few exist that use human user data collected live during gameplay. We will give some examples of recent notable ones. 

The AGAIN dataset~\cite{melhart2022arousal} features 37 hours of annotated gameplay footage across 9 different games created for the dataset. These 9 custom games were all heavily influenced by off-the-shelf games, and though they were created for this dataset, the purpose was not to evoke one particular emotion, but rather observe player affect. It is heavily documented and readily accessible after the submission of a terms of use agreement. The FUN\textit{ii} database~\cite{beaudoin2019funii} contains physiological data from 190 players during gameplay of Ubisoft's Assassin's Creed Unity~\cite{ac_unity} and Assassin's Creed Syndicate~\cite{ac_syndicate}. The first dataset of its kind in games, it features datatypes such as electrocardiography, electrodermal activity, and electromyography. The associated publication is heavily documented however access to the dataset was not available at the time of writing due to a broken link. The platformer experience dataset~\cite{karpouzis2015platformer} uses a facsimile of the original Mario Bros. game and is largely behavioural, providing detail about the participant's behaviour and game preferences and demographics, however also includes pre-processed data regarding the participant's facial features during gameplay. It is freely accessible through a link to a Google Drive folder. 

When focusing on VR studies in particular, there is a notable low transparency regarding data analysis and its availability across the field, with an example~\cite{Lanier2019Virtual} discussed in \ref{subsecFree}. In many cases, even when there is an effort to publish the data used, it is often restricted to the data "that support the findings of this study" \cite{yildirim2019cybersickness, banakou2016virtual}, with no apparent way to access the entirety of the data collected for further investigation.

 \section{Measurements}
\label{measure}
As well as ensuring construct validity of the study, the choice of tools to measure the target variables of interest is relevant to the replicability of the study. In order for the database to be used among other researchers, the tools should be relevant to the study, but the source material for these tools should also be available. That is, if there are a number of self-report questionnaires implemented, it would benefit researchers to have access to the questions used. This applies both to existing measurement tools that are adopted for the study, but also any new questions or questionnaires implemented in the study. From a dataset creation perspective, this might look like including both the raw data from the questionnaire, as well as any synthesis of the questionnaire subsets, and including the questions (and scoring instructions) with the dissemination of the research, or in the data repository. Often, questionnaires will be adapted from other fields, however these adapted questions are not accessible without directing contacting the researcher. These should also be made available within the dissemination of the work, or if there are limits to the quantity of detail permitted, a link to a permanent repository with any other supplementary material. 

For hardware that measures target variables (such as heart rate sensors often using in affective computing research), there may be less flexibility or freedom in which equipment to use due to cost and availability. There is often a range in the quality and accessibility of some physiological hardware (for example, NeuroSky's Mindwave\footnote{https://store.neurosky.com/pages/mindwave} compared to Brain Product's ActiCHamp\footnote{https://www.brainproducts.com/solutions/actichamp/}), however some aspects of these devices can be considered. For example, there are devices and their companion software that are completely open-source, giving full transparency to the researcher about the inner workings of the device. These devices are preferred as they enable easy collaboration between researchers through re-sharing of any modifications (mods) or additions to the experiment. Some recent examples include OpenBCI's Ultracortex electroencephalography (EEG) headset\footnote{https://shop.openbci.com/products/ultracortex-mark-iv} and the Emotibit\footnote{https://www.emotibit.com}, which measure a variety of physiological data (heart rate, electrical activity of skin, etc.), as well as open-source facial recognition software such as DeepFace\footnote{https://github.com/serengil/deepface} and OpenFace\footnote{https://github.com/TadasBaltrusaitis/OpenFace}. 
To assist the researcher, we have provided some questions here and throughout to ask as the experiment is being designed in order to consider transparency and replicability practices:
\begin{itemize}[noitemsep,nolistsep]
    \item If using a well established measurement, has it been validated or used in this context before?
    \item Do researchers have to pay or is it low-cost to access this measurement?
    \item Can I share this tool (such as a full set of questionnaire questions in the Appendix) or is there something else I can use that is more accessible?
    \item I am making my own questionnaire, how can I ensure that readers have access to the full set of questions, scoring information, and equipment information?
\end{itemize}

\section{Game Choice}
\label{game}
Regarding game choice, similar principles as measurement choice applies, albeit with a few additional considerations. When choosing a game for the study, the first decision that must be made is whether an off-the-shelf game will be used, or if a custom game will be created for the purpose of the study. Both options have benefits and disadvantages, and neither option is free from transparency issues. 

\textbf{Custom.}
As highlighted by~\cite{jarvela2014practical}, custom games created by the researcher have a number of benefits, including the ability to control all experimental details and potential confounding variables. However, custom games introduce a number of issues with bias as they are specifically designed to address the research question of interest. Often, the custom games are not made publicly available, limiting the ability for other researchers to replicate the same study. This discourages any further work being built upon the experimental design, but also reduces the transparency of the actual task itself. Textual descriptions will often be provided, but are not always adequately descriptive. As such, readers may be unaware of any bias in the design or implementation of the task that may favour the preferred outcomes of the study. Thus we make the suggestion that when a custom game is chosen or required for their study, the source code be made available on a repository, both for transparency of the minute details of the experiment, but to also make it accessible for researchers to replicate the study. 

\textbf{Off-the-shelf.}
Going beyond the choice of game validity purposes, there are additional considerations for the use of off-the-shelf games that may help or hinder transparency and replicability. Off-the-shelf games are an excellent choice in terms of both the generalisability of study findings, and, depending on the game, the ability to modify the experimental environment for better control. In some cases, games have strong modification (known commonly as modding) systems that allow nearly all aspects of the game to be changed, analysed or logged. For example, a modding API for Minecraft~\cite{minecraft_software_2011} has been developed and maintained since it's beta release, whereas the Creation Kit~\cite{bethesda_softworks_create_2012} has allowed for modding of most games in the Elder Scrolls Series, as well as many other games made by Bethesda. With purchase of these games, the software to modify these games is free, making it accessible for researchers to use off-the-shelf games in their research. 

Similar to custom games, researchers may not be able to give detailed enough textual descriptions of the game for replications, however basic details of the off-the-shelf games are likely already available if the game is accessible to the public. That being said, any modifications that are made to off-the-shelf games for the experimental design must be made aware in describing the experimental procedure. Additionally, platforms exist that allow people to upload their modifications to the game; these should be utilised by researcher for easy distribution and transparency on the inner workings of the experiment. Minecraft~\cite{minecraft_software_2011}, for example, mainly uses CurseForge\footnote{https://www.curseforge.com/} (this also the point of access for players to download and implement these modifications for enhanced gameplay), and Bethesda games such as Skyrim~\cite{bethesda_softworks_elder_2011} uses NexusMods\footnote{https://www.nexusmods.com/}. A core limitation of using games with modification support is that it does narrow down the number and type of games available to use, however the trade-off of increased transparency and replicability outweighs this limitation. Moreover, user-generated content and game modifications are increasing in popularity, with a number of newer games having support for adding or modifying game content; Garry's Mod~\cite{garrysmod_software_2006} and Roblox~\cite{roblox_software_2006} are examples of games where the purpose is for players to create their own games within it using the Lua programming language. To conclude, some researcher questions to consider during experimental design:
\begin{itemize}[noitemsep,nolistsep]
    \item I'm using an off-the-shelf game, where can I upload the modifications I have made to the game for others to access?
    \item I'm using a custom game, will I be able to make this publicly available on a repository? If not, is there a game I could use that I can make available or is already publicly available?
\end{itemize}


\section{Procedure}
\label{procedure}

\subsection{Methodology Considerations}
This section will detail some examples of best practice methodologies that prevent discrepancy between an original study and a replication study. Not all of these will be applicable to every game, but they are recommendations to consider in implementing the game task and event coding.

Regardless of the game, in the majority of game studies, players will need to be given instructions on how to interact with the task or control a character. These instructions need to be scripted in order to ensure reliability between participants, as no one participant should receive more or less information about the task than another. However, a script spoken by the researcher is still prone to human errors and any factors regarding the researcher's voice that may impact the participant's understanding. Therefore we recommend the use of text on the screen (this is common in human user research generally~\cite{whitley2012principles}), or, to improve immersion of the game, audio or textual (ideally both) instructions provided by either a character in game, or a narrator (provided by a text-to-voice algorithm, for example). For games where it is not possible to integrate this directly, one may also use a pseudo narrator where the audio speech is played over top of the game. The audio files or source code that integrates the speech should also be provided in documentation and distribution of the experiment detail and data. Providing the instruction script will also allow for easier alternate-language replication studies as it will not rely on the researcher to transcribe the spoken audio for translating.

Regarding logging and event coding, it is important for the coding of game events to be precise and able to be calibrated with any other data collected during play. With a modification-friendly commercial game, this is often readily accessible, making it the gold standard for event coding as it can easily be implemented for replication studies. For a researcher creating a custom game, the full control of a custom game will ensure that a sufficient logger can be included. As with the rest of the data and methods, the source code for the logger should also be made available for data synchronisation and technical purposes. For games in which accessing a native or in-built logging system is not possible, alternatives exist. As suggested in~\cite{jarvela2015stimulus}, keystroke and mouse-click logging could be included alongside a timestamped screen capture for post-processing. This is not ideal as it is both time-consuming and may be prone to human error if manually annotating the behaviours from the screen capture, which may be especially problematic for precise data analysis.

Internal validity directly relates to the ability for a study to be replicated~\cite{mohseni2015extensive}. In games and other digital research that uses software as the stimulus, the player can impact this internal validity by cheating or breaking the experiment. They can do so by accessing particular settings, or by killing or destroying particular non-player characters or items. In order to minimise this, the ability for the player to cheat needs to be removed. This is largely an issue for off-the-shelf games where natively, players may need to access particular settings during gameplay, however is not ideal for an experimental setting. Depending on the game, disabling these settings can be done internally in the development of a mod, or sometimes simply in the creation of a save file, which is the case for Minecraft~\cite{minecraft_software_2011}. In some cases, the player's character dying may also disrupt the experiment by respawning character somewhere else and outside of the experimental zone. This does not apply to all studies as characters dying is a natural aspect of a number of games, however when possible, it is recommended to prevent character death if it is not essential to the study.

\subsection{Documentation of Procedure}
Replicability studies are predicated on the ability to access the experimental detail from the original study~\cite{elliott2022taxonomy}, and therefore to promote replicability, researchers should aim to include as much information and source code as they are allowed. As given in the examples previously discussed, this includes details of all magnitudes, from the computer specifications to the exact survey questions asked of the participants. This is especially important in the creation of a public dataset to allow researchers to be fully informed about any potential cofounding variables when analysing the dataset and applying algorithms. Similarly, it is recommended to include all data and materials that were measured, even if they were not the focus of the study; this promotes transparency in the inclusion of findings that were not favourable towards the original research questions or hypotheses. It must be mentioned, however, that it is important to remove all personally identifying information from the provided dataset for the privacy of the participants. This should also be specified in the dataset or dissemination to communicate the reasons for any missing raw data~\cite{gebru2021datasheets}.

Often, page or word limit of research dissemination prevents all information from being easily shared with the research field, and in this case it is suggested that these details are uploaded to a repository or stable site. For datasets, this may simply be included in the GitHub or other secure repository alongside the data and mod or custom game source code. Often not included are the following: a complete step-by-step protocol, a list of hardware and software specifications, the complete bank of survey questions, and source code of a custom game or mod. These can all be uploaded securely and linked within the research dissemination for full transparency. Further, in some cases the dataset itself is given a separate DOI for easy citing and access.

\section{Proposed Case Study}
\label{casestudy}
In this section we detail the rationale and considerations in methodology design for an in-progress study that will also present a dataset upon completion. The case study demonstrates how to think about replicability and transparency in the decision-making process by giving real and practical descriptions. The actual decisions made by researchers when designing an experiment will differ with every study, and there are more considerations beyond what is discussed here, however the aim is to promote openness and transparency for the duration of the research process.

\subsection{Choice of game and task}
The aim of the upcoming experiment is to observe the relationship between various objective and subjective measures of cognitive load during natural gameplay to determine appropriateness of each measure in this context. The game chosen for this experiment is Minecraft, a 3D open-world procedurally generation sandbox game. There are no specific goals in Minecraft and thus a relevant task was created for the purpose of this experiment. This task involves the players constructing a house accompanied by a humanoid companion character that gives instructions, collects resources, and assists in building the player’s house. Though human companions are not native to the game, it is not unnatural for players to play with human characters as Minecraft has multiplayer functionality, and other types of companions (such as cats and horses) are native to the game. Minecraft is well known for its modding support from the developers and has a large community of hobbyist developers that create and share mods (including their source code) with the players. 

In this experiment, the Human Companions\footnote{https://www.curseforge.com/minecraft/mc-mods/human-companions} mod was expanded upon using the Forge API with Minecraft (Java Edition for PC) version 1.18.1 and Forge version 39.1.2. A number of behaviours were added to the companion, including the ability to find, remove, and collect wood blocks, and place those blocks onto a template provided in the game environment. A tutorial task was also included (prior to the main task) that requires players to complete a pseudo version of the main task in order to learn the controls of the game and become comfortable with the steps required to eventually complete the main task. All player instructions are pre-scripted and provided in-game via the chatbox and audio speech in order to ensure that the information provided is the same across every participant.

\subsection{Hardware}
\begin{figure}[h]
  \centering
  \includegraphics[keepaspectratio, height=12cm, width=9cm]{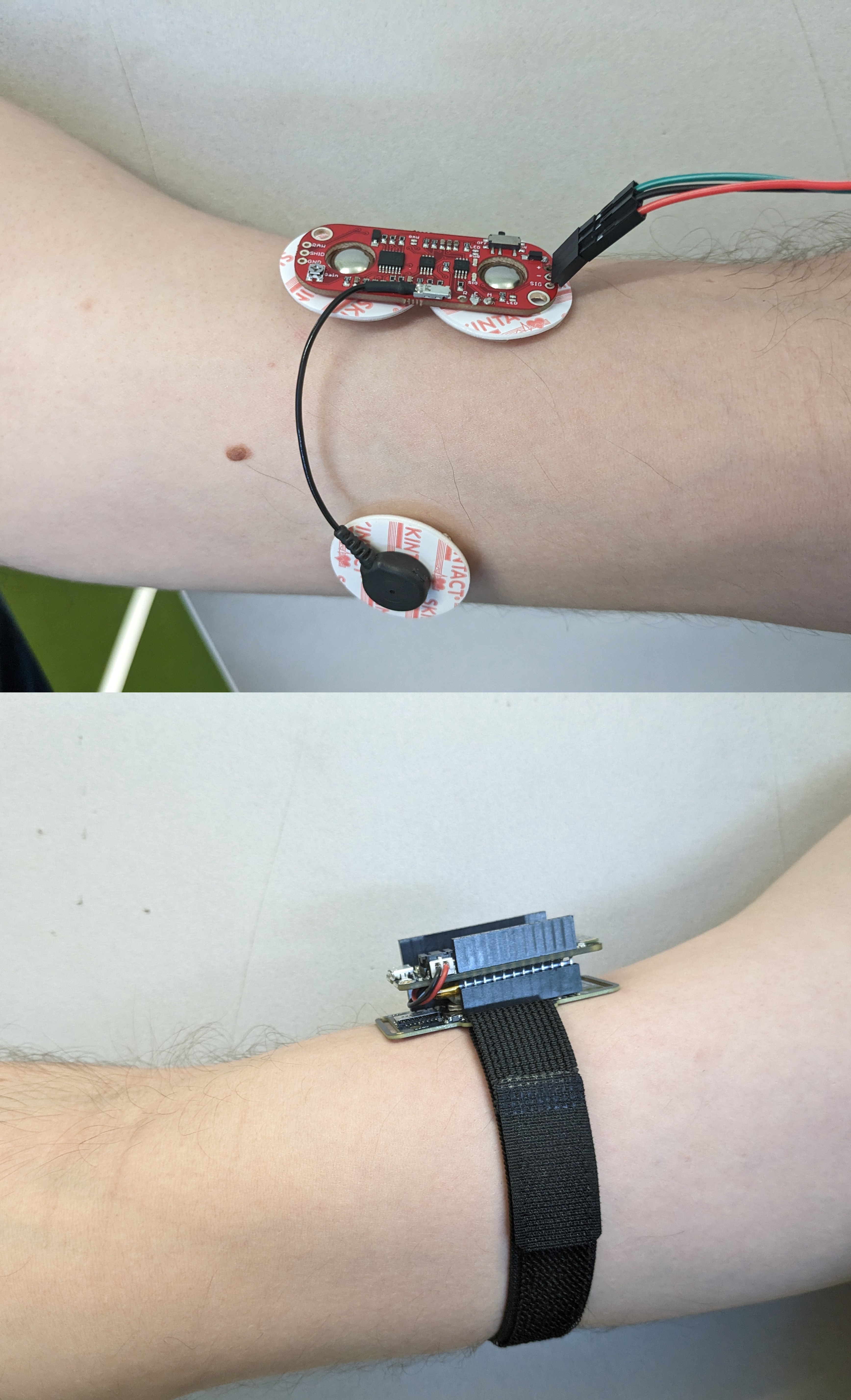}
  \caption{The Myoware Muscle Sensor attached to a participant's left bicep and the Emotibit attached to the right bicep.}
  \label{fig:emotibit}
  \end{figure}

The computer used to run the experiment and to collect data is an Alien Aurora R6 Desktop, it has an Intel Core i7-7700 CPU @ 3.60GHz Processor with 32GB of RAM and 64-bit operating system. It has Windows 10 Pro installed (version 21H2). The monitor is a BENQ GW2760 with a 60Hz refresh rate and 1920 x 1080 resolution. 
To collect the required physiological data, it was important that the devices were well-supported in the research field, and that there was long-term support for them from the developers. Open-source hardware and the accompanying software being open-source were priorities, and therefore the following devices were chosen:
\begin{itemize}[noitemsep]
    \item \textbf{OpenBCI Ultracortex Headset Mark IV.} A 16-channel dry EEG system will be set up using the 10:20 placement system. Both the main board (Cyton) and an extension board (Daisy) are required for the full 16 electrodes. The pro-assembled version in a size medium was purchased. A ~500 mAh lithium ion rechargeable battery is used to power the device, and the provided electrodes were replaced with 5mm comb reusable dry Ag-AgCl EEG electrodes (purchased from OpenBCI) for increased comfort and wearability. Data is transmitted to the computer using the Cyton Wi-Fi dongle.
    \item \textbf{Emotibit.} The Emotibit uses Adafruit Feather M0 WiFi board powered by a 400mAh Lithium ion battery and has 2 Ag/AgCL EDA electrodes for collecting a number of different measurements. It uses a microSD card to record the data and comes with straps for different placements. This is attached to the participants right bicep as shown in Figure~\ref{fig:emotibit}.
    \item \textbf{Myoware Muscle Sensor.} The Myoware Muscle Sensor v1.0 measures electromyography data and uses 3 embedded electrode connections. It is connected to an Arduino Uno v3 board, which is connected to an Adafruit 100ma Full-Speed USB Isolator (connected to the aforementioned PC) to ensure safety of the participants. This is attached to the participants left bicep as show in Figure~\ref{fig:emotibit}.
    \item \textbf{Logitech Camera.} A Logitech c922 Pro Stream Webcam records the video camera data of the participant's face. 
\end{itemize}

\subsection{Software}
To collect the data from the aforementioned sensors and other behavioural data, a number of different software are used:
\begin{itemize}[noitemsep]
    \item \textbf{OpenBCI GUI}. The open-source OpenBCI GUI (v5.1.0) is used to collect EEG data from the Ultracortex. The impedance is checked in the impedance widget to ensure a non-noisy signal (ideally under 500kOhms for dry electrodes). This data is sent to Lab Streaming Layer through the networking widget.
    \item \textbf{Lab Streaming Layer (LSL) and Lab Recorder.} To enable online processing and adaptation of game content in future experiments and time synchronization, the open-source Lab Streaming Layer program (v1.13.0) is used to transmit the data out of the GUI with no pre-processing into its accompanying data recording program, Lab Recorder.
    \item \textbf{Emotibit Oscilloscope and LSL marker stream.} This is the emotibit's open-source software that allows the streaming and recording of data. For local timestamps to be recorded, the oscilloscope (v1.5.10) can ingest a LSL marker stream ran with Python, this allows synchronisation of data streams offline.
    \item \textbf{Video camera capture.} A simple OpenCV (cv2 v4.7.0.68) Python script is used to initiate the video stream and print the timestamp of the first frame onto the video for later synchronisation of datastreams.
    \item \textbf{Minecraft Logging.} Embedded into the mod's code is an instance of Java's native logger that tracks and codes events within the game. It collects the timestamp (of the system's clock, with millisecond resolution) of each piece of companion dialogue (including instructions and progress messaging) and the time taken for the entire task. This same information (minus progress messaging) is also logged in the tutorial task.
\end{itemize}

\subsection{Questionnaires}
To collect subjective data regarding the participant's perceived cognitive load during the task, two questionnaires are administered post-task. Participants were also asked about their prior experience with Minecraft. Rationale are described below with the full list of questions in the Appendix:
\begin{itemize}[noitemsep]
    \item \textbf{Video Game Demand Scale (VGDS).} The VGDS is a subjective, video game specific measurement of mental load that explores four different factors of game demand; cognitive, emotional, physical, and social. Whilst not widely used, this has been chosen as the study aims to explore it's validity against other measures of cognitive load.
    \item \textbf{NASA-Task Load Index (NASA-TLX).} The NASA-TLX is a subjective questionnaire of task workload experience that contains six single-item subscales: mental demand. physical demand, temporal demand, performance, effort, and frustration. It is a widely used scale across a variety of domains, languages, and experimental tasks across the last 40 years and is considered the gold standard of measuring self-reported task workload. 
    \item \textbf{Prior experience questionnaire.} Participants will choose responses that best match their experiences with Minecraft and Minecraft-related products. This is collected in order to control and account for any impact on cognitive load that may occur as a result of prior experience. 
\end{itemize}

After the data collection is complete, all raw data (excluding the participant's raw facial data) will be uploaded to a github repository which will be accessible to interested researchers after agreeing to a terms of use form similar to~\cite{melhart2022arousal}. Alongside the data will be all questionnaire materials, the adapted Human Companions mod, and the code for implementing any data collection software. This persistent link will be available alongside the dissemination of research.

\section{Conclusion}

This paper demonstrated ways in which transparency, and consequently replicability, can be encouraged and implemented in human user video games research. A small number of human user video games datasets currently exist, and we hope that by providing some actionable advice and recommendations, we can promote researchers to share their protocols and data in order to further explore these phenomena and gain scientific consensus on aspects of gameplay. We recommended the choice of commercial games with strong modding support, open-source hardware and software in data collection, and sharing thorough instructions for replicating the study design. The paper is concluded with a proposed case study of these methods using an in-progress that aims to present a dataset alongside all available study details and decisions. The implementation of this work will not only improve overall transparency of games research, but also allow better collaboration in human games research by building scientific consensus.


\bibliographystyle{abbrv-doi}

\bibliography{bibliography}

\appendix

\begin{table*}[t]
\renewcommand{\arraystretch}{1.5}
\caption{Minecraft Experience Survey Questions}
\label{MinecraftExpQs}

\begin{center}
\begin{tabular}{|l|l|}
\hline
  \textbf{Question/s} & \textbf{Response Options} \\
  \hline
  \textbf{Q1} What best describes your experience with playing & “I play it daily” \\ 
  (the original) Minecraft? & “At a minimum, I play it once a week” \\ 
  & “At a minimum, I play it once a month” \\
  & “At a minimum, I play it once a year” \\
  & “I have at least played it once in the last 5 years” \\
  & “I have not played it in the last 5 years” \\
  & “Never played it” \\
  \hline
  \textbf{Q2} If you have played the original Minecraft game before, & “PC" \\ 
  what console/s have you played it on? & “Home Console (e.g. Xbox, Playstation, Wii U)" \\
  & “Handheld Console (e.g. Nintendo DS, Switch, Playstation Vita)"\\
  & “Mobile (e.g. Android, iPhone, iPad)"\\
  & “VR (e.g. Oculus, PSVR)" \\ 
  & “Other, please specify" \\
  \hline 
  \textbf{Q3} Which of the following Minecraft games have you played? & “Original Minecraft" \\
  & “Minecraft: Story Mode" \\
  & “Minecraft Dungeons" \\
  & “Minecraft Earth" \\
  & “Minecraft Education" \\
  & “Other, please specify" \\
  \hline
\end{tabular}
\end{center}
\end{table*}

\begin{table*}[t]
\renewcommand{\arraystretch}{1.5}
\caption{NASA-TLX Questions~\cite{hart1988development}}
\begin{center}
\begin{tabular}{|c|}
    \hline
    NASA-TLX questions \\
    \hline
    \textbf{Mental Demand} - How mentally demanding was the task? \\
    (1 = “Very \textbf{Low}"  7 = “Very \textbf{High}") \\
    \hline
    \textbf{Physical Demand} - How physically demanding was the task? \\
    (1 = “Very \textbf{Low}"  7 = “Very \textbf{High}") \\
    \hline
    \textbf{Temporal Demand} - How hurried or rushed was the pace of the task? \\
    (1 = “Very \textbf{Low}"  7 = “Very \textbf{High}") \\
    \hline
    \textbf{Performance} - How successful were you in accomplishing what you were asked to do? \\
    (1 = “\textbf{Perfect}"  7 = “\textbf{Failure}") \\
    \hline
    \textbf{Effort} - How hard did you have to work to accomplish your level of performance? \\
    (1 = “Very \textbf{Low}"  7 = “Very \textbf{High}") \\
    \hline
    \textbf{Frustration} - How insecure, discouraged, irritated, stressed, 
    and annoyed were you? \\
    (1 = “Very \textbf{Low}"  7 = “Very \textbf{High}") \\
    \hline
    \end{tabular}
    \end{center}
    \label{nasa-tlx}
\end{table*}

\begin{table*}[t]
\renewcommand{\arraystretch}{1.5}
\caption{VGDS Questions~\cite{bowman2018development}}
\begin{center}
\begin{tabular}{|c|c|c|c|c|c|c|}
    \hline
    \multicolumn{7}{|c|}{\textbf{Response Options}} \\
    \hline
    1 & 2 & 3 & 4 & 5 & 6 & 7\\
    Strongly Disagree & Disagree & Somewhat Disagree & Neither Agree nor Disagree &
    Somewhat Agree & Agree & Strongly Agree \\
    \end{tabular}
\begin{tabular}{|l|}
     \hline
     \textbf{Cognitive Demand} \\
     \hline
      \quad\quad The game was cognitively demanding. \\
      \quad\quad I had to think very hard when playing the game.\\
      \quad\quad The game required a lot of mental gymnastics.\\
      \quad\quad This game doesn’t require a lot of mental effort.\\
      \quad\quad The game made me draw on all of my mental resources.\\
      \quad\quad The mental challenges in this game had an impact on how I played.\\
      \quad\quad The game stimulated my brain.\\
      \hline
      \textbf{Emotional Demand} \\
      \hline
      \quad\quad The game tugged at my heartstrings. \\
      \quad\quad The game gave me the feels.\\
      \quad\quad I was moved by the game.\\
      \quad\quad I had a strong emotional bond with the game content.\\
      \quad\quad I had a lot of unexpected feelings during gameplay.\\
      \hline
      \textbf{Exertional Demand} \\
     \hline
      \quad\quad I was physically exhausted after playing. \\
      \quad\quad I felt strained after playing.\\
      \quad\quad My body felt drained after gameplay.\\
      \quad\quad The game was physically demanding.\\
      \hline
      \textbf{Social Demand} \\
      \quad\textit{Note}: Where we refer to “others" in some of these statements, this can refer to other avatars, characters, and/or players in the game.\\
      \hline
      \quad\quad Socializing was an important part of playing this game. \\
      \quad\quad While playing, I was aware of others in the game.\\
      \quad\quad I was compelled to interact with others in the game.\\
      \quad\quad I felt obligated to others, while playing.\\
      \quad\quad Being around others in the game had an impact on how I played.\\
      \quad\quad This game was socially demanding.\\
      \hline
\end{tabular}
    \end{center}
    \label{nasa-tlx}
\end{table*}

\end{document}